\documentclass[twocolumn,amsmath,amssymb,aps,pra,10pt,lettersize,superscriptaddress,showpacs]{revtex4-1}

\usepackage[dvips]{graphicx} 
\usepackage{braket}
\usepackage[usenames]{color}

\usepackage{ulem}

\graphicspath{{./Figure/}}

\definecolor{nblue}{rgb}{0.3,0.3,1.0}
\definecolor{ngreen}{rgb}{0.2,0.7,0.2}
\definecolor{nred}{rgb}{0.9,0.1,0}

\renewcommand{\prl}{Phys.\ Rev.\ Lett.\ }
\renewcommand{\pra}{Phys.\ Rev.\ A }
\renewcommand{\rmp}{Rev.\ Mod.\ Phys.\ }

\newcommand{\prx}{Phys.\ Rev.\ X }
\newcommand{\ncommun}{Nat.\ Commun.\ }
\newcommand{\PRA}[2]{\pra \textbf{#1}, #2 }
\newcommand{\PRL}[2]{\prl \textbf{#1}, #2 }

\newcommand{\articletitle}[1]{}

\DeclareMathOperator{\arctanh}{arctanh}
\DeclareMathOperator{\Tr}{Tr}

 \iftrue
  
  \newcommand{\delete}[1]{\textcolor{blue}{\sout{#1}}}
  \newcommand{\comment}[1]{\textcolor{blue}{\footnotesize #1}}
\else
  
  \newcommand{\delete}[1]{}
  \newcommand{\comment}[1]{}
\fi

\def\UTokyo{Department of Applied Physics, School of Engineering, The University of Tokyo,\\
7-3-1 Hongo, Bunkyo-ku, Tokyo 113-8656, Japan}
\def\UPalacky{Department of Optics, Palack\'y University, 17.\ listopadu 1192/12, 77146 Olomouc, Czech Republic}
\def\UMainz{Institute of Physics, Staudingerweg 7, Johannes Gutenberg-Universit\"{a}t Mainz, 55099 Mainz, Germany}

\begin{document}


\title{Noiseless conditional teleportation of a single photon}

\author{Maria Fuwa}
\affiliation{\UTokyo}
\author{Shunsuke Toba}
\affiliation{\UTokyo}
\author{Shuntaro Takeda}
\affiliation{\UTokyo}
\author{Petr Marek}
\affiliation{\UPalacky}
\author{Ladislav Mi\v{s}ta,~Jr.}
\affiliation{\UPalacky}
\author{Radim Filip}
\affiliation{\UPalacky}
\author{Peter van Loock}
\affiliation{\UMainz}
\author{Jun-ichi Yoshikawa}
\affiliation{\UTokyo}
\author{Akira Furusawa}
\email{akiraf@ap.t.u-tokyo.ac.jp}
\affiliation{\UTokyo}

\date{\today}

\begin{abstract}
We experimentally demonstrate the noiseless teleportation of a
single photon by conditioning on quadrature Bell measurement
results near the origin in phase space and
thereby circumventing the photon loss that otherwise occurs
even in optimal gain-tuned continuous-variable quantum teleportation. 
In general, thanks to this loss suppression, the noiseless conditional teleportation
can preserve the negativity of the Wigner function for 
an arbitrary pure input state and 
an arbitrary pure entangled resource state.
In our experiment, the positive value of the Wigner function at the
origin for the unconditional output state, $W(0,0)=0.015 \pm 0.001$,
becomes clearly negative after conditioning,
$W(0,0)=-0.025 \pm 0.005$, illustrating
the advantage of noiseless conditional teleportation.
\end{abstract}



\maketitle


Quantum teleportation \cite{PRL70-1895, PRA49-1473, PRL80-869} is
one of the key quantum information protocols.
It corresponds to the process of transferring an unknown quantum state
between two spatially separated parties by means of an entangled
state and classical communication. The original quantum-state teleportation
has been generalized to scenarios where either the entangled state
(gate teleportation \cite{Nature402-390, PRL90-117901}) or
the measurement basis (one-way quantum information
processing \cite{PRL86-5188, PRL97-110501}) is modified.
As a consequence, teleportation plays an important role
in general quantum information schemes such as quantum key distribution
in complex networks \cite{Madsen2012} and
fault-tolerant quantum computation \cite{PRA71-032318, Rev79-135}.
Teleportation is also being considered as the main method of
building quantum interfaces between various physical
realizations of a harmonic oscillator
\cite{Hammerer2010,Hoffer2011,Barzanjeh2012}.

Recently there have been attempts to reconcile the two main approaches
to quantum teleportation,
discrete-variable (DV) and continuous-variable (CV) teleportation,
in order to combine the strengths and mitigate the weaknesses of either approach.
This culminated in an experimental demonstration of deterministic high-fidelity
CV teleportation of DV quantum states \cite{Takeda.nature2013}.
In that experiment, near-optimal gain tuning was employed in the feed-forward loop,
thus suppressing the addition of thermal photons and resulting in an almost pure attenuation of the
teleported state --- an effect typically less harmful to the nonclassical features of a quantum state.
For example, when teleporting a single photon state, pure attenuation only leads to an extra vacuum term.
In a deterministic teleportation scheme, however, this additional vacuum is unavoidable in order
to ensure preservation of the fundamental commutation relations.
If a measurement is performed in the photon number basis, which is typically the case
in DV optical quantum information processing, the occurrence of an extra vacuum term usually only
affects the success probability, i.e., the efficiency, but not the fidelity of the quantum protocol.
On the other hand, in most CV and also in hybrid CV-DV applications, the vacuum contribution is highly undesirable.
For instance, it may render the Wigner function of a single photon state positive,
even though the negativity of the Wigner function is an important quantum feature necessary in many applications \cite{PRL81-5932, PRL102-120501, PRL88-097904, Mari2012}.

In this letter, we experimentally test the final refinement of the teleportation protocol --- noiseless conditional teleportation.
In this protocol, we turn the attenuation in the gain-tuned teleportation into 
an almost noiseless attenuation with no additional vacuum term.
Noiseless attenuation generally still alters a quantum state, but it can preserve the purity and nonclassical properties of pure states \cite{Micuda.prl2012,supplement1}.
In teleportation, it can be gradually approached by conditioning dependent on the outcome of the CV Bell-state measurement (CV-BSM) on the sender's side.
We demonstrate this conditioned noiseless teleportation protocol for a single photon state \cite{Mista.pra2010}, which could be subject to other quantum operations
in a larger DV protocol before it enters and after it leaves the teleporter.
Our main figure of merit will be the value of the Wigner function at the phase-space origin, $W(0,0)$.
This value, when negative, is a very sensitive manifestation of nonclassical quantum features necessary for the most advanced quantum protocols \cite{PRL81-5932, PRL102-120501, PRL88-097904, Mari2012, Mista.pra2010}.

\begin{figure}[!t]
\begin{center}
\includegraphics[width= \linewidth, clip]{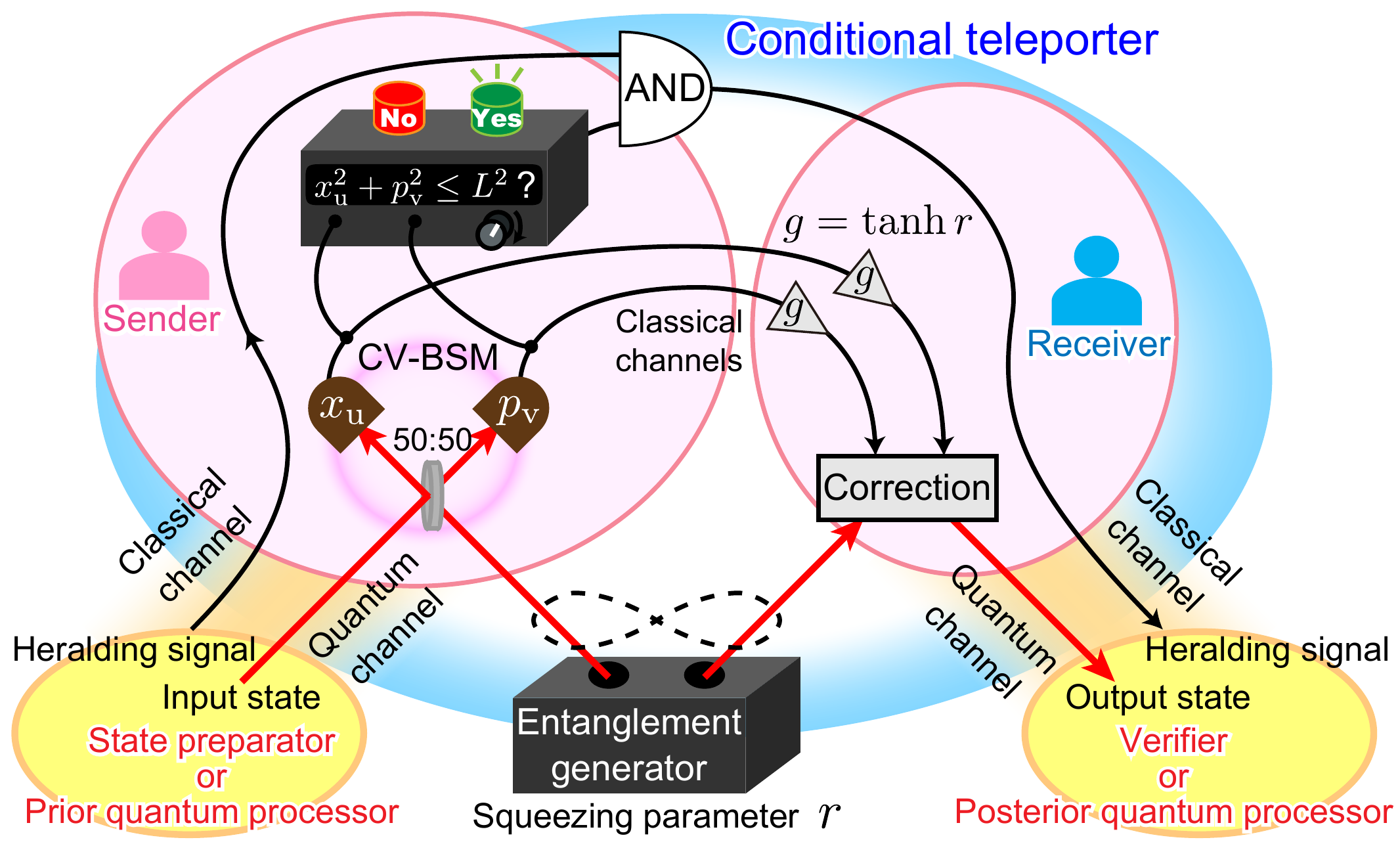}
\caption{
Overview of gain-tuned continuous-variable quantum teleportation conditioned by the sender.
} \label{fig:scheme_overview}
\end{center}
\end{figure}

An overview of our scheme is given in
Fig.~\ref{fig:scheme_overview}. In the prototypical CV
teleportation, parties A and B first share a CV entangled state,
ideally the Einstein-Podolsky-Rosen (EPR) state
$\sum_{n=0}^{\infty}\tanh^n r \ket{n}_A\ket{n}_B$. This state,
which has perfect photon-number correlations for
any amount of entanglement, can be generated by combining two
orthogonally squeezed states on a balanced beam splitter
\cite{PRA40-913}. The amount of the CV entanglement is
proportional to the squeezing parameter $r$
\cite{Vidal_02}. In the infinite energy limit of $r
\rightarrow \infty$, both quadratures (amplitude and phase or
position and momentum) of the state are also perfectly correlated.

The CV teleportation is now started by combining the input state
$\ket{\psi}$ with one part of the EPR state on a balanced beam
splitter and subjecting the resulting modes to a pair of homodyne
measurements, which yields the values of quadrature operators $\hat{x}$ and
$\hat{p}$ (CV-BSM). The CV-BSM results $(x_{\mathrm{u}},p_{\mathrm{v}})$
are multiplied by feed-forward gain $g$ and used to perform a
correction via displacement on the remaining part of the EPR state
in order to obtain a replica of the input state. This imperfect replica generally
contains a certain amount of thermal noise --- a consequence of the finite
entanglement --- but when the gain is tuned to $g = \tanh r$, the
CV teleportation becomes equivalent to a purely attenuating
channel described by \cite{supplement5}
\begin{equation}\label{eq:attenuation_channel}
\ket{\psi} \rightarrow \sum_{k=0}^{\infty} \frac{1}{k! \sinh^{2k} r}\hat{a}^k \left[(\tanh r)^{\hat{n}}\ket{\psi}\bra{\psi}(\tanh r)^{\hat{n}}\right] \hat{a}^{\dag k},
\end{equation}
where $\hat{a}$ and $\hat{n}$ are the annihilation and photon number
operators, respectively. The attenuating process in
\eqref{eq:attenuation_channel} consists of two different
sequential contributions. The first one is a \textit{noiseless
attenuation}, represented by operator $(\tanh r)^{\hat{n}}$, and it is
accompanied by random energy annihilation, represented by a
mixture of operators $\hat{a}^k$. The latter is responsible for
the reduction of the state's purity and
the disappearance of any negative
values of the Wigner function. 
In the case of a perfect single photon
state $\ket{1}$ at the input, the output state becomes
$\tanh^2 r \ket{1}\bra{1} + (1- \tanh^2 r ) \ket{0}\bra{0}$, where
the vacuum term originates solely from the
random annihilation process. For this state, the negativity of
the Wigner function vanishes when the entanglement of the shared
state is below a certain bound, $r\leq\arctanh(1/\sqrt{2})$.


The quality of the teleported state can be substantially improved by conditioning on the data obtained from the CV-BSM prior to the feed-forward.
By accepting only those events, for which the measured results $x_{\mathrm{u}}$ and $p_{\mathrm{v}}$ satisfy $x_{\mathrm{u}}^2 + p_{\mathrm{v}}^2 \leq L^2$, we can suppress the noise that is caused by the random measurement outcomes and feed-forward.
The conditioning can be understood as insertion of a filter in a classical channel that transmits heralding signals, as
illustrated in Fig.~\ref{fig:scheme_overview}.
In the limit of $L\rightarrow 0$, we can achieve perfect noiseless teleportation,
\begin{equation}\label{eq:noiseless}
\ket{\psi} \rightarrow (\tanh r)^{\hat{n}}\ket{\psi},
\end{equation}
only resulting in noiseless attenuation without the detrimental
effect of the random annihilation described by
\eqref{eq:attenuation_channel}. Clearly, a single
photon state is faithfully teleported, and likewise
any Fock state is transferred with unit fidelity for any
nonzero entanglement. Importantly, unit
fidelity is also obtained for dual-rail qubits
$\alpha\ket{0,1}+\beta\ket{1,0}$ \cite{Takeda.nature2013}
by utilizing two identical
conditional teleporters \cite{supplement3}. 
The improvement through conditioning is very well observable by looking at the achievable
Wigner-function negativity of the teleported single photon state
\cite{Mista.pra2010, supplement1}.

For more general input states, in particular states
with an unfixed photon number \cite{footnote}, the remaining noiseless
attenuation could be locally compensated after the teleportation
by a noiseless conditional amplification approximately applying the
operator $\gamma^{\hat{n}}$, where the gain $\gamma=1/\tanh r$ is
inversely proportional to the amount of entanglement used in the
teleportation. The noiseless amplification can be based on various
concepts \cite{Ralph2009,Marek2010}, which were already
experimentally tested
\cite{Usuga2010,Zavatta2010,Xiang2010,Ferreyrol2010,Koscis2013,Chrzanowski2014}.
The compensation of the noiseless attenuation has been
experimentally verified for coherent states \cite{Muller.pra2012}
and qubits \cite{Micuda.prl2012}. Importantly, the noiseless
amplification cannot compensate the attenuation
\eqref{eq:attenuation_channel}, but only the noiseless attenuation
\eqref{eq:noiseless}. In the presence of impure resources, the
teleportation is no longer completely noiseless, but it still
outperforms all other forms of CV teleportation
\cite{supplement2}.


Our CV teleporter relies on the deterministic one described in
Refs.~\cite{Takeda.nature2013,Lee.science2011}, while our
probabilistic but heralded preparation of the input single photon
state is described in more detail in Ref.~\cite{Takeda.pra2013}.
The source laser is a continuous-wave Ti:sapphire laser with a
wavelength of $860$ nm (the frequency is denoted by $\omega_0$).
As for the preparation of the input state, a nondegenerate
optical parametric oscillator (NOPO) containing a PPKTP crystal
(type-0 quasi-phase-matched) is weakly pumped by a
frequency-shifted second harmonic ($2\omega_0 + \Delta \omega$),
probabilistically creating signal and idler photon pairs
($\omega_0$ and $\omega_0 + \Delta \omega$, respectively), where
the frequency separation ($\Delta \omega$) of $590$ MHz is the
free spectrum range (FSR) of the NOPO\null. Then, detection of an
idler photon by a silicon avalanche photodiode heralds a
signal photon, whose wavepacket corresponds to the linewidth of
the NOPO modes, $6.2$ MHz of half-width at half maximum (HWHM).
The average creation rate is $7800$ /s with a pump power of $3$
mW. As for the teleportation, the EPR state is created by
combining two squeezed vacuum states at a balanced beam splitter.
Each squeezed vacuum state is created from a degenerate optical
parametric oscillator (OPO) containing a PPKTP crystal, pumped by
$125$ mW of the second harmonic ($2 \omega_0$). The OPOs (FSR $1$
GHz) are made smaller than the NOPO, in order to make the
bandwidth of the squeezed vacuum states (HWHM $12.9$ MHz) wider
than that of the single photon wavepacket. The resulting EPR state
corresponds to the squeezing parameter $r=1.62 \pm 0.03$ and
the effective loss $l=0.20 \pm 0.02$. Homodyne detectors and electric
amplifiers in the feed-forward loop have bandwidths over $10$ MHz,
which is sufficiently wide to teleport the input single photon
wavepackets.

\begin{figure}[!b]
\begin{center}
\includegraphics[width=\linewidth , clip]{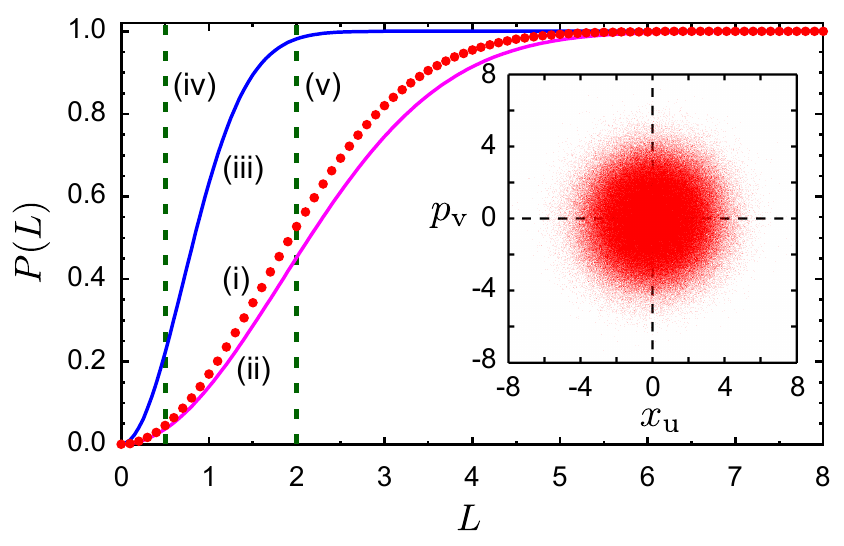}
\caption{
Probability $P(L)$ of CV-BSM falling inside conditioning radius $L\ge0$ ($\hbar = 1$).
(i)~Experimental results.
(ii)~Theoretical prediction based on the experimental input photon number distribution together with the single-mode
squeezing parameter $r=1.62$ for the EPR state and optical loss $l=0.20$.
(iii)~Theoretical curve for CV-BSM on two-mode vacuum states.
(iv,v)~$L=0.5$ and $L=2.0$ used in Fig.~\ref{fig:results_tomography}.
Inset: experimentally obtained CV-BSM distribution.
} \label{fig:results_probability}
\end{center}
\end{figure}

To confirm our strategy, quadratures of the output states are measured by a homodyne detector with a local oscillator (LO) phase scanned, and the outcomes are stored for $640,000$ events together with the LO phase information and the outcomes of the CV-BSM.
Then, the conditional output states are reconstructed by quantum tomography \cite{JOptB6-6} collecting only those events satisfying $x_{\mathrm{u}}^2 + p_{\mathrm{v}}^2 \leq L^2 $.
The success probability $P(L)$ of the conditioning for various $L\ge0$ is shown in Fig.~\ref{fig:results_probability}.
The deviation of the experimental probability from the theoretical prediction may be attributed to our simplified model including the assumption of a symmetric EPR state or the drift of parameters during the experiment. 
On the other hand, the input state is reconstructed by eight-port homodyne tomography \cite{Takeda.pra2013} with $100,000$ events.
Error bars for the reconstruction are estimated by using the bootstrap method \cite{Efron.Book(1994)}.

\begin{figure*}[!t]
\begin{center}
\includegraphics[width=\linewidth , clip]{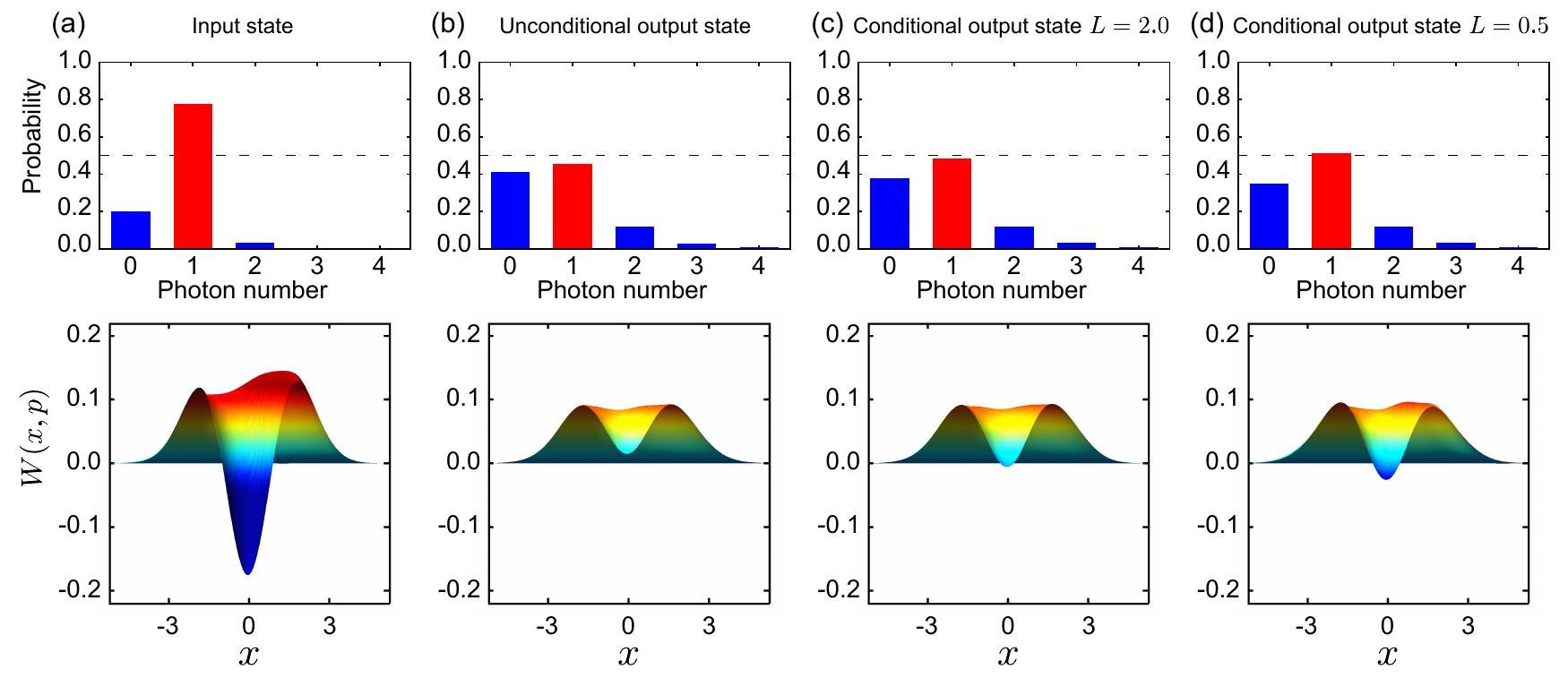}
\end{center}
\caption{
Experimental quantum states reconstructed by the maximum likelihood method~\cite{JOptB6-6}.
(a)~Input single-photon state.
(b)~Output state without conditioning.
(c,d)~Output states conditioned by $L = 2.0$ and $L = 0.5$, respectively.
Top: photon number distribution.
The sum of odd photon number components exceeding the horizontal line at $0.5$ means a Wigner function negative at the origin.
Bottom: side view of the Wigner function.
} \label{fig:results_tomography}
\end{figure*}


The experimental results are shown in Fig.~\ref{fig:results_tomography}, where we can see the photon number distributions and Wigner functions for the input state and for three output states differing by the level of conditioning.
The initial input state in Fig.~\ref{fig:results_tomography}(a) is a statistical mixture of single photons ($p_1 = 0.772 \pm 0.007$), vacuum ($p_0 = 0.195 \pm 0.004$) and higher photon numbers ($p_{n \geq 2} = 0.033 \pm 0.003$), where $p_n$ is the photon number probability.  The origin of the Wigner function of the input state has a value of $W(0,0) = (1/\pi)\sum_{n=0}^\infty (p_{2n} - p_{2n+1}) = -0.174 \pm 0.005$, which is roughly one half of the maximum possible value ($-1/\pi\approx-0.318$) and a clear indication of nonclassicality.

When the input state is teleported with unity gain, $g = 1.0$, a negative Wigner function with $W(0,0) = -0.023 \pm 0.004$ is obtained, but this only happens at the expense of increasing the higher photon number contributions, which are typically unwanted in DV qubit-type experiments. 
These can then be suppressed by gain tuning, but only with an extra vacuum contamination \cite{Takeda.nature2013, PRA88-042327}. As a result, the Wigner-function negativity will be reduced.
Figure~\ref{fig:results_tomography}(b) shows the unconditional output state with a tuned feed-forward gain $g=0.89$, a value close to $\tanh r = 0.93$.
Although there is still a certain level of discrepancy from the attenuation channel in \eqref{eq:attenuation_channel} due to the impure EPR state, multiphoton components are decreased compared to the unity-gain case. 
However, excess vacuum contamination renders the Wigner function positive with $W(0,0) = 0.015 \pm 0.001$.
To improve the negativity again, the method of conditioning is an effective means, as will be demonstrated in the following.

Figure~\ref{fig:results_tomography}(c) shows the results of
moderate conditioning with $L = 2.0$, $P(L) = 0.53$. Now the
output state has a negative Wigner function in the
origin, $W(0,0) = -0.006 \pm 0.002$, ascertaining strong nonclassicality.
This enhancement comes from the elimination of the vacuum contribution 
and, at the same time, an increase of the single photon contribution from $p_1 = 0.449 \pm 0.001$ to $p_1 = 0.480 \pm 0.002$.
Thus, with the little sacrifice of dropping the teleportation rate from $100\%$ to $53\%$, the quality of the output state can be enhanced to a strongly nonclassical state with a negative Wigner function.

When the conditioning radius is further narrowed to $L=0.5$, $P(L) = 0.05$, the vacuum elimination is further enhanced, as shown in Fig.~\ref{fig:results_tomography}(d).
The single-photon contribution rises to $p_1 = 0.511 \pm 0.007$, resulting in a further growth of the dip in the Wigner function to have a negative value of $W(0,0) = -0.025 \pm 0.005$.
This value is comparable to that of $g=1.0$, thereby demonstrating that the combination of gain-tuning and conditioning yields better results than unity-gain teleportaion.

All the results above show the strength of our conditioning method.
The degree of conditioning can be tuned to meet various requirements: a negative Wigner function was obtained with success rate of $53\%$ and the negativity matches that in the unity-gain teleportation, while significantly suppressing the higher photon numbers, with success rate of $5\%$.
In order to approach ideal noiseless attenuation, the purity of the EPR state must be improved, for which the effective loss $l$, arising mainly from propagation loss, mode mismatch in homodyne measurements, and phase fluctuations in interferometers, need to be reduced.

Conditional quantum teleportation is a very versatile tool.
Apart from its suitability for the transfer of both CV and DV \cite{Takeda.nature2013, PRA88-042327, supplement3} states of light, it is capable of conditionally implementing a broad class of \textit{Gaussian and non-Gaussian quantum nonlinear filters} on arbitrary unknown states of traveling light beam \cite{supplement4}.

The probabilistic teleportation can be also used as a \textit{light-matter interface} for atoms and mechanical oscillators \cite{Hammerer2010,Hoffer2011}.
The purity-preserving aspect of the operation with conditional Bell measurement allows to perfectly transfer individual Fock states, produced by efficient single photon guns \cite{Rempe.nphys2007,Yoshikawa.prx2014}, from optical modes to modes of atomic ensembles or mechanical oscillators.
This interface requires only weak Gaussian entanglement, which can be produced by a weak light-matter interaction of the down-conversion kind, simply accessible for both the atomic ensembles and mechanical oscillators \cite{Hammerer2010,Hoffer2011}.


In conclusion, we have experimentally demonstrated noiseless conditional teleportation of a single photon state.
By conditioning on the results of the Bell-type measurement before applying the feed-forward loop, we were able to reduce the value of the Wigner function at the phase-space origin from positive values without conditioning to negative values after conditioning.
This clearly confirms the feasibility of the noiseless teleportation, which can be used to qualitatively enhance the transmission of quantum states.
Our results represent an important step for the noiseless realization of quantum relays, quantum repeaters, quantum memories and interfaces, as well as possibly even linear-optics quantum computing.


We acknowledge support from the SCOPE program of the MIC of Japan, PDIS, GIA, G-COE, and APSA commissioned by the MEXT of Japan, FIRST initiated by the CSTP of Japan, ASCR-JSPS, the Australian Research Council Centre of Excellence CE110001027, 
grant P205/12/0577 of Czech Science Foundation 
and Czech-Japan bilateral project LH13248 of the Ministry of Education, Youth and Sports of Czech Republic.
M.F.\ and S.T.\ acknowledges financial support from ALPS.

\makeatletter
\balancelastpage@sw{%
  \onecolumngrid
 }{}%
\newpage
\twocolumngrid
\makeatother

\onecolumngrid

\begin{center}\Large\bf Supplemental Material\end{center}

\section{Preservation of negativity of Wigner function}
The asymptotic noiseless teleportation applied to Gaussian coherent states only decreases the amplitude $\ket{\alpha}\rightarrow \exp(-(1-g^2)|\alpha|^2/2)\ket{g\alpha}$, where $g=\tanh r$ is the transmissivity of the noiseless attenuation.
Any pure superposition state $\ket{\psi}=\int C(\alpha)\ket{\alpha} d^2 \alpha$, written in the over-complete basis of coherent states, approaches after the teleportation a state of the form $\int \exp(-(1-g^2)|\alpha|^2/2) C(\alpha)|g\alpha\rangle d^2 \alpha$ up to normalization.
For pure Gaussian states, $C(\alpha)$ is a complex Gaussian function and asymptotically, the teleported state remains Gaussian.
On the other hand, pure non-Gaussian states have a non-Gaussian wave function and equivalently, $C(\alpha)$ is also non-Gaussian.
All pure non-Gaussian states exhibit a negativity of the Wigner function, as a result of Hudson's theorem \cite{Hudson}.
Since the noiseless teleportation asymptotically approaches the transformation $\ket{\alpha}\rightarrow \exp(-(1-g^2)|\alpha|^2/2)\ket{g\alpha}$ of Gaussian coherent states, noiseless teleportation preserves the non-Gaussianity of pure input states.
Due to Hudson's theorem, the output pure non-Gaussian state produced by the noiseless teleportation then also exhibits negativity of the Wigner function for any amount of the preshared entanglement.
This is an important and generic feature of the noiseless teleportation, which makes it distinct from the unconditional gain-tuned teleportation implementing an attenuating channel, where the negativity of the Wigner function can be lost.

\section{Realistic resources}
\textit{Noiseless teleportation of a realistic single photon.} ---
For mixed input states, the noiseless teleportation based on
finite entanglement (finite $r$) will not always preserve the
negativity of the Wigner function. For any experimental tests it
is therefore advantageous to use input states with high purity. As
an example of the input state, let us consider an
attenuated single photon state
$\eta\ket{1}\bra{1}+(1-\eta)\ket{0}\bra{0}$, which has no
negativity of the Wigner function for $\eta\leq 0.5$. The
noiselessly teleported state has the form $[\eta\tanh^2
r\ket{1}\bra{1}+(1-\eta)\ket{0}\bra{0}]/[1-\eta(1-\tanh^2 r)]$,
which has negativity of the Wigner function if $\tanh^2
r>(1-\eta)/\eta$. It can be now seen that in order to faithfully teleport a
small negativity of the Wigner function, a higher amount of
entanglement is required. And, conversely, if the purity of the single
photon state is high ($\eta\rightarrow 1$), the negativity remains
even for weak entanglement $(r\ll 1)$. It should be noted that for
the same input state, the deterministic teleportation with
optimal gain tuning produces $\eta\tanh^2 r\ket{1}\bra{1}+(1-\eta\tanh^2
r)\ket{0}\bra{0}$, where negativity appears only if $\tanh^2
r>1/(2\eta)$ such that the entanglement must exceed a
certain finite bound even for an ideal single
photon. The conditional teleportation is therefore the ideal
device for transferring quantum states in the presence of weak
shared entanglement.

\vspace{\baselineskip}

\textit{Conditional teleportation with noisy entangled state.} ---
A general noisy entangled resource state can be expressed as a
mixture of states $ \hat{A}_k \otimes \hat{B}_l\sum_{m}g^m
\ket{m}_A\ket{m}_B$, where $\ket{m}_A\ket{m}_B$ stand for Fock
states in modes A and B and $\hat{A}_k$ and $\hat{B}_l$ are Kraus operators
representing independent noisy channels. Using the noiseless
conditional teleportation, the input state  $|\psi_{in}\rangle$ is
transformed into a mixture of states $\hat{B}_l g^{\hat{n}} \hat{A}_k^{T}
\ket{\psi_{in}}$. We can then understand the noiseless
teleportation as a perfect teleportation with additional steps
taking effect in the following order: noise introduced to mode A,
noiseless attenuation originating from the finite entanglement,
and noise introduced to mode B. This is the bare minimum of noise
which is introduced by any form of CV teleportation. The
conditional teleportation achieves this regime, because unlike the
other forms of teleportation, it is not concerned with
averaging over all the possible measurement results of the CV-BSM. We
can therefore say that the conditional teleportation exploits the
preshared entanglement optimally with respect to the quality of the
teleported state.

\section{Noiseless teleportation of entanglement}
The noiseless teleportation is especially useful for the conditional transfer of hybrid entangled states,
\begin{align}
\frac{\left(\ket{\Psi}_C\ket{\Phi}_A-\ket{\Phi}_C\ket{\Psi}_A\right)}{\sqrt{2-2|\braket{\Psi|\Phi}|^2}},
\end{align}
which contain $1$ ebit of entanglement for arbitrary different
states $\ket{\Psi}_i = \sum_n c_n \ket{n}_i$ and
$\ket{\Phi}_i= \sum_n d_n \ket{n}_i$, $i=A,C$. Note that,
although here we are using a rather general expression without
supposing orthogonality between the states $\ket{\Psi}_i$ and
$\ket{\Phi}_i$, the Gram-Schmidt orthonormalization
$\ket{\Phi}_i=\ket{\Psi}_i\braket{\Psi|\Phi} +
\ket{\Psi_\perp}_i\sqrt{1-|\braket{\Psi|\Phi}|^2}$ with
$\braket{\Psi|\Psi_\perp}=0$ can be utilized to check that the
above states exactly contain $1$ ebit. When both sides of the
state undergo the noiseless teleportation, the state is
transformed into
$\ket{\Psi'}_C\ket{\Phi'}_A-\ket{\Phi'}_C\ket{\Psi'}_A$, where
$\ket{\Psi'}_i=\sum_n c_n g^n \ket{n}_i/\sqrt{\sum_n |c_n|^2
g^{2n}}$ and $\ket{\Phi'}_i=\sum_n d_n g^n \ket{n}_i/\sqrt{\sum_n
|d_n|^2 g^{2n}}$. However, even though the state is different, it
still contains $1$ ebit of entanglement for arbitrarily low value
of $g$. In this way, the noiseless teleportation can perfectly
transfer fragile hybrid entangled states, such as entangled large
coherent-state superpositions or entangled high-number Fock states -- something that 
unity gain and gain-tuned, unconditional teleportation protocols cannot.

\section{Noiseless teleportation as a filter}
The specific form of the filter is programmed by using a particular quantum state $\ket{f}_{AB}\propto \sum_{k,\ell} f_{k\ell}\ket{k}_A\ket{\ell}_B$ in place of a generic two-mode squeezed vacuum state.
For sufficiently small conditioning radius $L$, the modified teleportation implements the filter $\hat{F}=\sum_{k,\ell} f_{k\ell}\ket{k}\bra{\ell}$, which transforms an arbitrary input state $\hat{\rho}$ into $\hat{\rho}'=\hat{F}\hat{\rho}\hat{F}^{\dagger}/(\Tr \hat{F}\hat{\rho}\hat{F}^{\dagger})$.
If the program state $\ket{f}_{AB}$ is Gaussian, we can implement a class of Gaussian filters described by $\hat{F}=\hat{S} g^{\hat{n}}\hat{S}'$ \cite{Fiurasek2013}, where $0 \leq g\leq 1$ and $\hat{S},\hat{S}'$ are squeezing operations.
Alternatively, when the program state is from a low-dimensional Hilbert space, such as a two-qubit state, we can implement quantum scissors \cite{Lvovsky2003}.
We can also change the program state into $\ket{f}_{AB}\propto \sum_{k=0}^{N} g^{k} \ket{kk}_{AB}$, where $g \geq 1$ in order to approach a more compact version of the noiseless quantum amplifier \cite{Ralph2009} when $N$ increases.
From this perspective, the presented results already serve as experimental verification of the most basic filter $\hat{F}=g^{\hat{n}}$, where $0 \leq g\leq 1$, corresponding to the noiseless attenuation of the single-photon test state.
Preservation and improvement of negativity of the Wigner function is an important witness for the high quality of this filter operation.

\section{Mathematical supplement}

Equation~(1) in the main text, corresponding to an attenuation channel, is equivalent to the standard way of expressing 
an amplitude damping channel using Kraus operators,
\begin{subequations}
\begin{gather}
\hat{\rho} \to \sum_{k=0}^{\infty} \hat{A}_k \hat{\rho} \hat{A}_k^\dagger, \\
\hat{A}_k = \sum_{\ell=k}^{\infty} \sqrt{\binom{\ell}{k}}\sqrt{(1-\gamma)^{\ell-k}\gamma^k}\ket{\ell-k}\bra{\ell}, \label{eq:random_annihiration}
\end{gather}
\end{subequations}
where $1-\gamma=\tanh^2r$.
Equation~\eqref{eq:random_annihiration} is in accordance with the probability of losing $k$ photons from $\ell$ photons by a linear optical loss under which each photon is randomly lost with a probability $\gamma$.

The CV-BSM is represented by a set of joint projective operators of the form
\begin{align}
\hat{\Pi}_{i,j}(x_{\mathrm{u}},p_{\mathrm{v}}) = \hat{D}_i(\sqrt{2}x_{\mathrm{u}},\sqrt{2}p_{\mathrm{v}})\ket{\text{EPR}}_{i,j}\bra{\text{EPR}}\hat{D}_i^\dagger(\sqrt{2}x_{\mathrm{u}},\sqrt{2}p_{\mathrm{v}})
\end{align}
where $\ket{\text{EPR}}_{i,j}=\sum_{k=0}^{\infty} \ket{k}_i\ket{k}_j$ is an ideal EPR state ($r\to\infty$) without normalization, and $\hat{D}_i(x_{\mathrm{u}},p_{\mathrm{v}})=\exp(-i(x_{\mathrm{u}}\hat{p}_i-p_{\mathrm{v}}\hat{x}_i))$ is a phase-space displacement operator.
The factor $\sqrt{2}$ is introduced in order to compensate the weight factors from the balanced beam splitter.
The conditioning means collecting only those projection events satisfying $x_{\mathrm{u}}^2 + p_{\mathrm{v}}^2 \leq L^2$.
The limit $L\to 0$ leads to a projection onto an ideal EPR state $\hat{\Pi}_{i,j}(0,0)=\ket{\text{EPR}}_{i,j}\bra{\text{EPR}}$, which purifies the teleportation process 
from noise according to
\begin{align}
\ket{\psi}_V \to
\left(\sum_k\null_{V,A}\bra{k,k}\right)\left(\ket{\psi}_V\sum_\ell g^\ell\ket{\ell,\ell}_{A,B}\right)
=g^{\hat{n}}\ket{\psi}_B,
\end{align}
corresponding to a noiseless attenuation with $g=\tanh r$ determined by the resource two-mode squeezing level.


\begin{thebibliography}{99}
\bibitem{footnote}
Note that only for a coherent-state
input, $\ket{\alpha}$, even the deterministic attenuation channel obtainable
via optimally gain-tuned, unconditional teleportation leads
to a noiseless attenuation, $\ket{(\tanh r) \alpha}$. In other words, it is not useful
to apply noiseless conditional teleportation upon coherent states.
\end{thebibliography}
\end{document}